# Atomically-thin metallic Si and Ge allotropes with high Fermi velocities

Chin-En Hsu[1], Yung-Ting Lee[2], Chieh-Chun Wang[1], Chang-Yu Lin[1], Yukiko Yamada-Takamura[3], Taisuke Ozaki[4], and Chi-Cheng Lee[1,5]*

[1]Department of Physics, Tamkang University; Tamsui, New Taipei 251301, Taiwan.

[2]Department of Electrophysics, National Yang Ming Chiao Tung University; Hsinchu, Taiwan.

[3]School of Materials Science, Japan Advanced Institute of Science and Technology (JAIST); 1-1 Asahidai, Nomi, Ishikawa 923-1292, Japan.

[4]Institute for Solid State Physics, The University of Tokyo; 5-1-5 Kashiwanoha, Kashiwa, Chiba 277-8581, Japan.

[5]Research Center for X-ray Science, College of Science, Tamkang University; Tamsui, New Taipei 251301, Taiwan.

*Corresponding author. Email: cclee@mail.tku.edu.tw

**Abstract:** Silicon and germanium are the well-known materials used to manufacture electronic devices for the integrated circuits but they themselves are not considered as promising options for interconnecting the devices due to their semiconducting nature. We have discovered that both Si and Ge atoms can form unexpected metallic monolayer structures which are more stable than the extensively studied semimetallic silicene and germanene, respectively. More importantly, the newly discovered two-dimensional allotropes of Si and Ge have Fermi velocities superior to the Dirac fermions in graphene, indicating that the metal wires needed in the silicon-based integrated circuits can be made of Si atom itself without incompatibility, allowing for all-silicon-based integrated circuits.

Ever since graphene was discovered (*1*), the physical properties in two-dimensional (2D) materials have been extensively explored for realizing higher-performance electronic devices used in modern technology (*2*, *3*). Heterostructures with diverse electronic structures can be further built as playing with "Lego blocks" by stacking layered materials with different twisted angles (*4*, *5*). Among the known 2D materials, silicene and germanene are of great interest owing to their hexagonal structures that can host Dirac fermions, akin to graphene (*6*, *7*). Another fact is that the hexagonal structure forms the Lego layer for building the 3D diamond structure along the (111) direction, which gives the silicon and germanium used in the semiconductor industry. A metallic silicene-based layer useful for devices has also been experimentally demonstrated (*8*).

The miniaturization of integrated circuits requires not only high-performance transistors but also highly efficient interconnects to keep up with Moore's law (*9*, *10*). Due to the semiconducting nature, Si and Ge allotropes have not generally been considered as a good option for connecting the miniaturized silicon-based transistors. Instead, the interconnecting wires are usually made of metals such as copper; however, diffusion of copper into semiconducting materials could severely affect the designed doping effects (*11*, *12*). Other metals might possess insufficient conductivity or lack availability for manufacturing. For ultimate miniaturization, it is desirable to have an atomic-scale interconnect made of Si atom itself with high conductivity, allowing for all-Si-based integrated circuits.

We have discovered atomically-thin metallic layers composed of Si and Ge atoms with high Fermi velocities superior to the Dirac fermions in graphene from first-principles calculations, and they are more stable than silicene and germanene, the well-known thinnest 2D Si and Ge allotropes, respectively. The strategy for finding this structure is to reduce the thickness of the diamond structure along the [001] direction and study underexplored 2D structures having a square lattice. Note that the direction is aligned with the Si(001) substrate surface normal, which is commonly adopted for fabricating a variety of devices. As shown in Fig. 1A, the 1D buckling chain, like the polyacetylene described by the Su-Schrieffer-Heeger (SSH) model (*13*), can be extracted from the (001) surface. After introducing new bonding between chains, a bridge-like structure (Bridge) is formed. By further introducing the interchain buckling in Bridge structure, the buckled re-bonded diamond(001) layer is revealed and will be dubbed as D(001).

To demonstrate D(001) is both thermodynamically and dynamically stable, we have calculated the total energy and phonon dispersion using the OpenMX code (*14*) by adopting two sophisticated approximations within the density functional theory (*15*), the local density approximation (LDA) and generalized gradient approximation (GGA) (*16*, *17*). To confirm the stability, we have also performed the calculations using another first-principles package, Quantum Espresso (*18*, *19*), and arrived at the same conclusion [see supplementary materials for the computational details]. As listed in Table 1, the Si and Ge allotropes in Bridge structure have already possessed lower total energies than the low-buckled and high-buckled silicene and germanene, respectively. The total energies in the so-called "$MoS_2$" structure (*20*) are also compared. After introducing the interchain buckling, the total energies of Si D(001) and Ge D(001) become the lowest among all the studied structures listed in Table 1, respectively. However, this is not the case for C D(001) where no interchain buckling can be identified, that is, C D(001) may be reduced to Bridge structure. This is consistent with the fact that puckered structures are preferred by Si and Ge atoms, allowing for diverse forms of bonding (*20, 21*).

Although Si D(001) and Ge D(001) are energetically preferred, the D(001) structure might not be dynamically stable. Recall that high-buckled silicene and germanene also possess lower total energies than their low-buckled forms but they are in fact not stable due to the presence of imaginary-frequency vibrational modes (*6*). To demonstrate the dynamical stability of Si D(001) and Ge D(001), the phonon dispersions of Si D(001) and Ge D(001) within both LDA and GGA are presented in Figs. 1B to E. Overall, the frequencies in Ge D(001) are lower than those in Si D(001) due to the heavier mass of Ge atom, and the LDA calculations give higher frequencies than the GGA ones due to the delivered shorter lattice constants that may lead to overbinding [see supplementary materials for the lattice parameters]. The absence of imaginary frequencies in all the dispersions advocates the dynamical stability of Si D(001) and Ge D(001).

The electronic band structures and density of states shown in Figs. 2A to D indicate that both Si D(001) and Ge D(001) are metallic, similar to the case of 2D metallic borophene in contrast to the semiconducting bulk boron allotropes (*22*). Note that a high Fermi velocity of $1.09 \times 10^6$ m/s can be identified within GGA after arranging borophene on the Al(111) surface (*23*). The electronic velocities obtained from the calculated momentum matrix elements (*24*), which are also reflected by the slopes in the band dispersion, are presented in Figs. 2E and F. Like the metals that are proposed to replace copper for the interconnects (*25*), the electronic velocities in

Si D(001) and Ge D(001) reach the order of $10^6$ m/s. The high velocities can be visualized from the pyramids shown in Figs. 2E and F with steep slopes in both x and y directions.

In Figs. 2G and H, the Fermi surfaces generated using FermiSurfer software (*26*) are presented and the highest Fermi velocities are both found at Y in the 2D Brillouin zone. The magnitudes of the velocities at Y reach $1.88\times10^6$ ($1.89\times10^6$) and $1.82\times10^6$ ($1.89\times10^6$) m/s in Si D(001) and Ge D(001), respectively, within GGA (LDA). Graphene hosts Dirac fermions with the Fermi velocity around $0.9\times10^6$ m/s within LDA (*27, 28*). As shown in Figs. 2I and J, the magnitudes of velocities along the Y-to-Γ direction in both the Si D(001) and Ge D(001) cases are prominently higher than those along the K-to-Γ direction in graphene within either LDA or GGA. Since the calculated Fermi velocities in graphene can be modified by introducing a substrate (*27*) and many-body interactions beyond LDA or GGA (*28, 29*), the Fermi velocities in Si D(001) and Ge D(001) could be higher.

The bands near the Fermi level at Y exhibit an interesting feature. While the one-dimensional Dirac fermions with the remarkably high velocities are revealed along the y direction, the bands along the perpendicular direction are relatively flat and form a four-fold-degenerate nodal line, as shown in Figs. 2E and F. After taking spin-orbit coupling into account, the nodal line is split into two two-fold degenerate nodal lines, as shown in Figs. 2I and J. But the four-fold degeneracy at Y is intact, forming a Dirac point. The Dirac point at Y, one of the time-reversal-invariant momenta, is protected by both time-reversal symmetry and inversion symmetry. The degeneracy is highly tunable, for example, by breaking the inversion symmetry with adatoms.

Si and Ge atoms crystallizing in the D(001) structure with high stability is actually unexpected. Each Si or Ge atom has to accommodate its two *s*- and two *p*-orbital electrons to six anisotropic bonds surrounding the atomic center. Consequently, the strong covalent bonds such as the fully filled $\sigma$ and $\pi$ bonds in silicene and germanene cannot be formed and the metallicity with partially filled *p* orbitals is anticipated. As shown in Figs. 2A and B, the high velocities along the x and y directions near the Fermi level are contributed from the $p_x$ and $p_y$ orbitals, respectively. By considering that the D(001) layer is composed of two buckled rectangular-lattice layers, the steep slopes can be attributed to the strong intralayer $p_x$-$p_x$ and $p_y$-$p_y$ hopping strengths. The charge density distribution presented in Fig. 3, which is generated using VESTA software (*30*), further reveals that the occupied orbitals form a close-packed structure in the *ac* plane, as a result of rehybridization of s, $p_x$, and $p_z$ orbitals. On the other hand, electrons distribute more along the y direction in the buckled rectangular-lattice layer, suggesting that the bonds formed by the $p_y$ orbitals are closer to covalent bonding.

The revealed D(001) structure has demonstrated again that the bonding in the Si and Ge allotropes is quite flexible, and more low-dimensional Si and Ge structures with multiple buckling forms are expected to possess competitive or even lower energies, especially given that the diamond structure sets the lowest-energy limit among the Lego blocks. The atomically-thin metallic Si D(001) and Ge D(001) with high Fermi velocities also advocate the possibility of fabricating all-Si-based electronic devices that can be used in the semiconductor industry, ranging from transistors to interconnects. The buckling forms, thickness, orientation, and doped elements all play key roles in the miniature technology. These findings motivate more studies on the Si and Ge phases in between 2D D(001) and 3D diamond structures under various conditions, such as uniaxial strain and/or temperature.

**Table 1. Total energy of monolayer structures composed of C, Si, and Ge atoms.** Relative total energies of high-buckled (HB), Bridge, D(001), and diamond structures composed of C, Si, and Ge atoms to the total energies of graphene, low-buckled (LB) silicene, and LB germanene, respectively. The energy in the "$MoS_2$" structure is also compared. Both LDA and GGA results calculated using OpenMX and Quantum Espresso codes are listed. The lattice parameters can be found in the supplementary materials. C Bridge structure is equivalent to C D(001) structure.

| TOTAL ENERGY (eV/atom) | | OpenMX | | Quantum Espresso | |
|---|---|---|---|---|---|
| Element | Structure | LDA | GGA | LDA | GGA |
| C | Graphene | 0 | 0 | 0 | 0 |
| | Bridge/D(001) | 1.700 | 1.769 | 1.731 | 1.786 |
| | Diamond | -0.085 | 0.100 | -0.037 | 0.137 |
| Si | LB silicene | 0 | 0 | 0 | 0 |
| | HB silicene | -0.018 | 0.102 | -0.044 | 0.066 |
| | Bridge | -0.235 | -0.119 | -0.227 | -0.114 |
| | D(001) | -0.243 | -0.127 | -0.240 | -0.127 |
| | $MoS_2$ | -0.095 | -0.031 | -0.078 | -0.016 |
| | Diamond | -0.812 | -0.698 | -0.755 | -0.641 |
| Ge | LB germanene | 0 | 0 | 0 | 0 |
| | HB germanene | -0.204 | -0.127 | -0.146 | -0.075 |
| | Bridge | -0.251 | -0.148 | -0.195 | -0.099 |
| | D(001) | -0.273 | -0.176 | -0.216 | -0.125 |
| | $MoS_2$ | -0.193 | -0.141 | -0.158 | -0.115 |
| | Diamond | -0.674 | -0.538 | -0.621 | -0.484 |

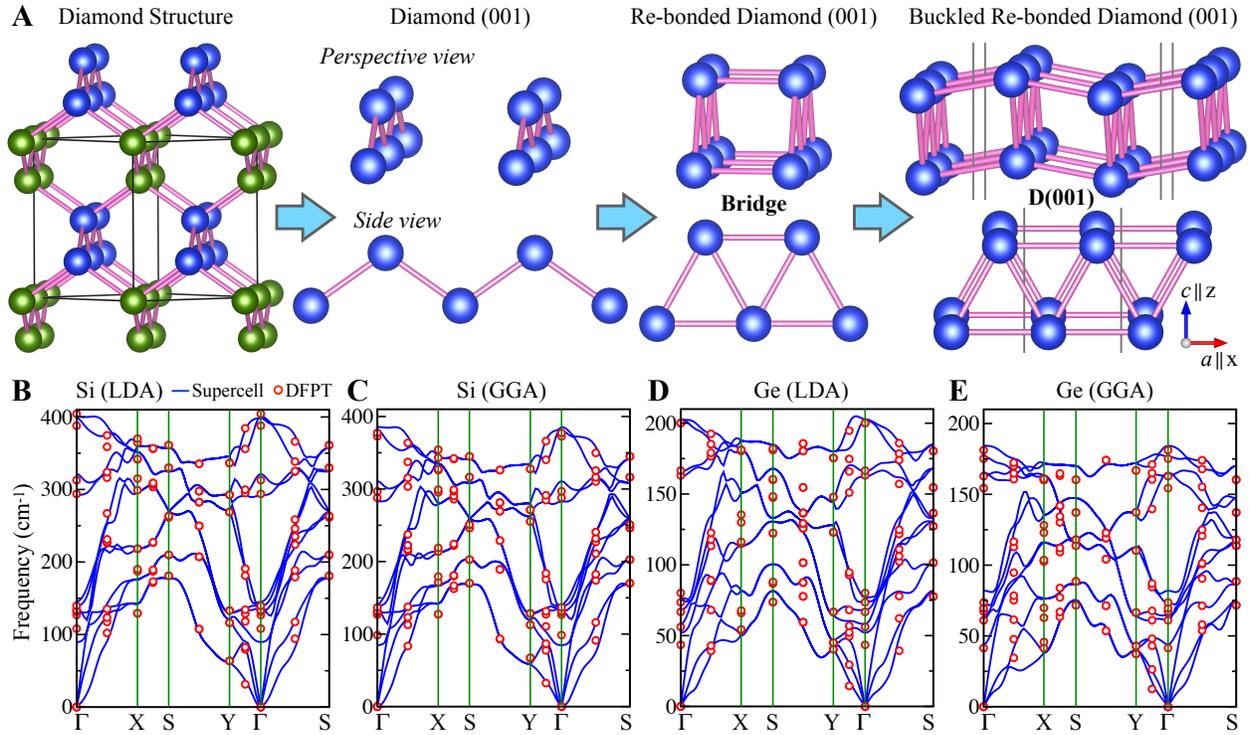

**Fig. 1. Structure and phonon dispersion of Si D(001) and Ge D(001).** (**A**) Schematics of the formation of Bridge and D(001) structures from the diamond structure. Phonon dispersion of Si D(001) within (**B**) LDA and (**C**) GGA. Phonon dispersion of Ge D(001) within (**D**) LDA and (**E**) GGA. Γ, X, S, and Y denote (0, 0), (0.5, 0), (0.5, 0.5), and (0, 0.5) in units of the reciprocal lattice vectors. The curves are obtained from the supercell force-constant calculations using OpenMX code and the circles denote the result of density functional perturbation theory using Quantum Espresso code. The structures shown in (**A**) are generated using VESTA software.

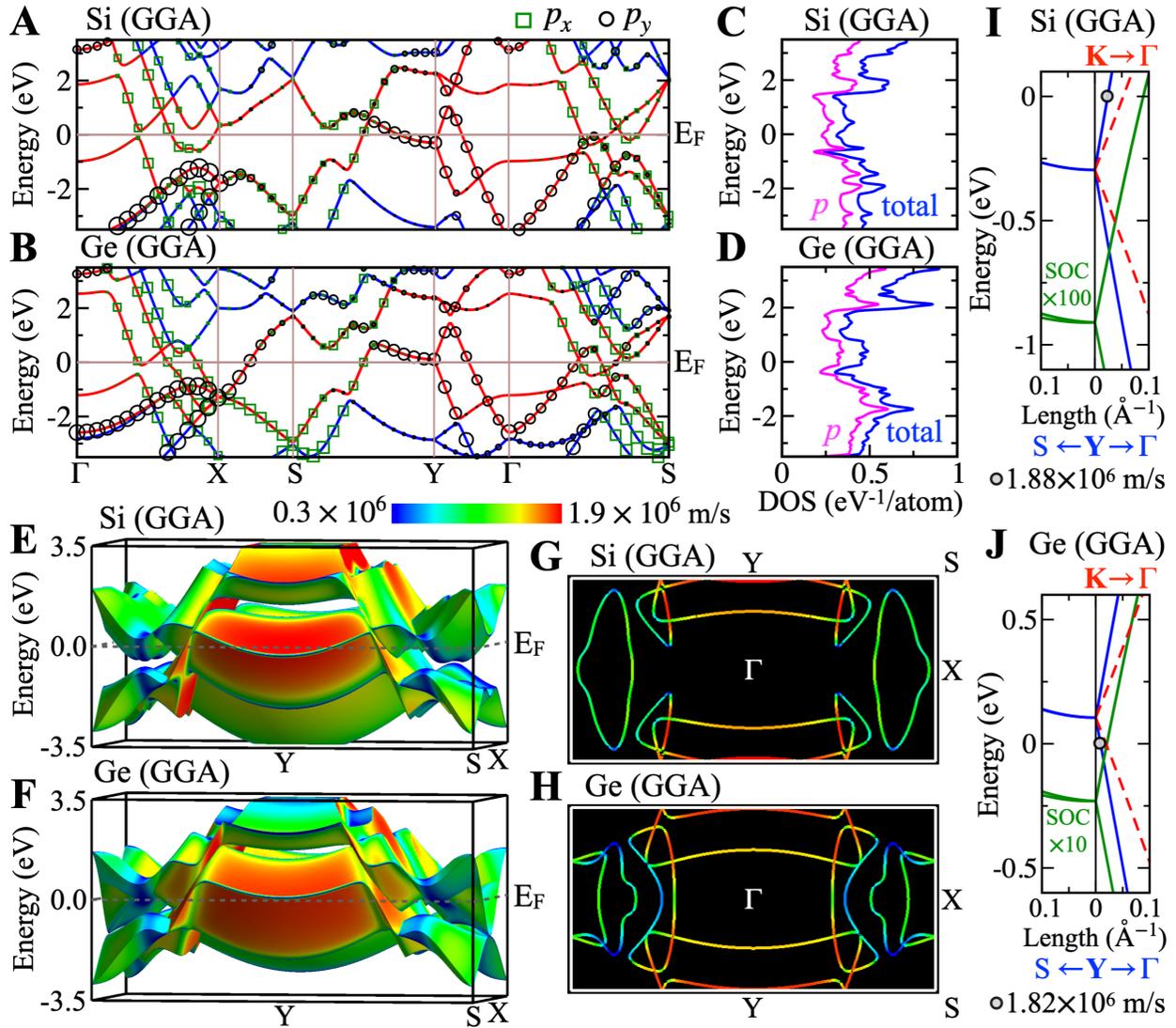

Fig. 2. Electronic band structures and Fermi surfaces of Si D(001) and Ge D(001). Band structures of (**A**) Si D(001) and (**B**) Ge D(001) within GGA using OpenMX code. Density of states with the *p*-orbital contribution of (**C**) Si D(001) and (**D**) Ge D(001). The red bands in (**A**) and (**B**) are colored with the magnitudes of velocities (red: high, blue: low) for (**E**) Si D(001) and (**F**) Ge D(001). $E_F$ denotes the Fermi level. Fermi surfaces of (**G**) Si D(001) and (**H**) Ge D(001). After taking spin-orbit coupling into account, the Dirac cone in graphene (red dashed curves) is aligned to the Dirac cones in (**I**) Si D(001) and (**J**) Ge D(001) (blue curves). The bands with the strength of spin-orbit coupling enlarged by (**I**) 100 times and (**J**) 10 times are presented by green curves. The velocities at the Fermi levels as marked by the circles in (**I**) and (**J**) are also listed. The plots in (**E** to **H**) are generated using FermiSurfer software. The LDA result and the band structures obtained using Quantum Espresso code are shown in the supplementary materials.

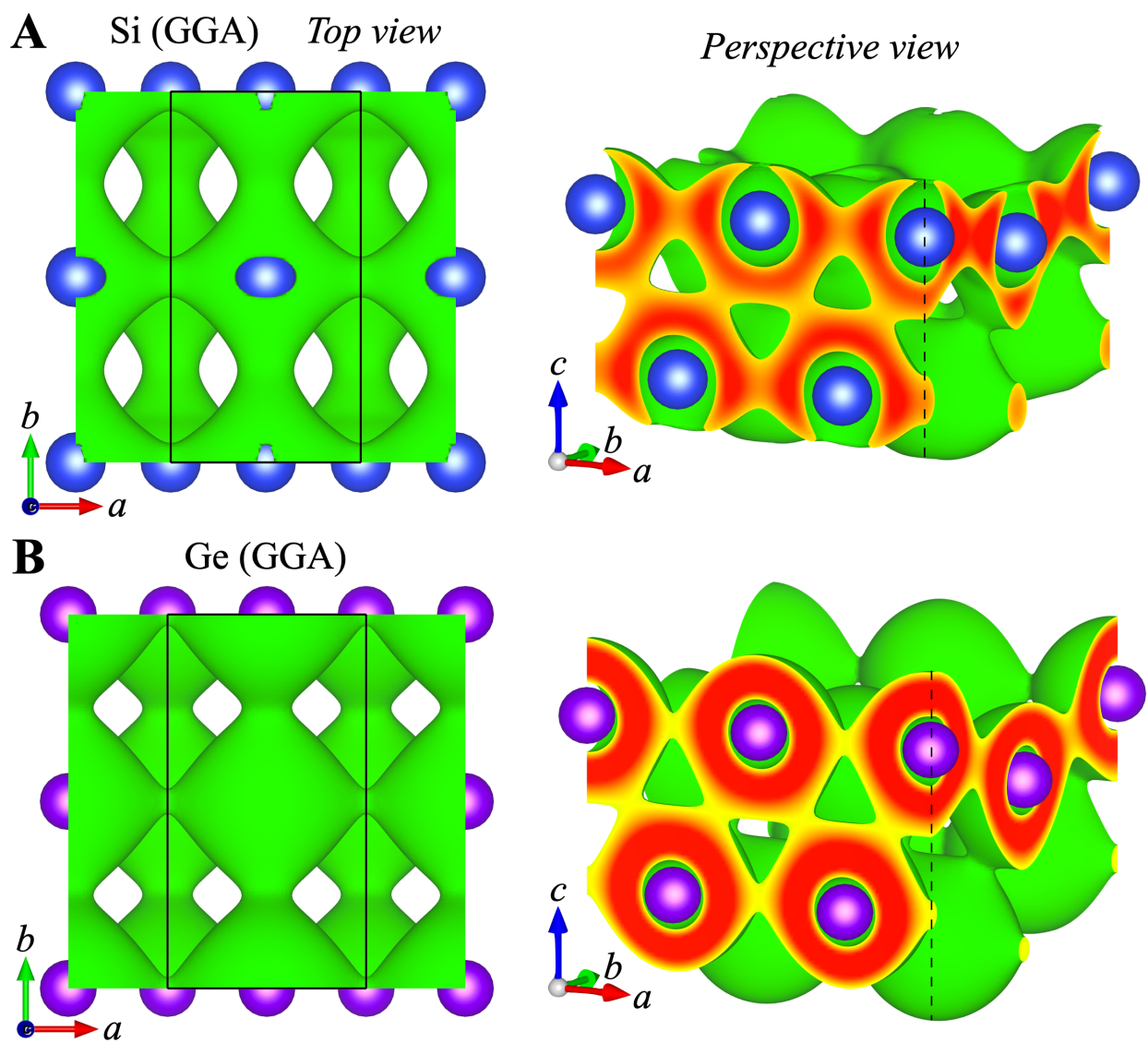

**Fig. 3. Charge density distribution in Si D(001) and Ge D(001).** Top view and perspective view of isosurface of charge density at (**A**) 0.052 electrons/bohr$^3$ in Si D(001) and (**B**) 0.035 electrons/bohr$^3$ in Ge D(001) within GGA. The result within LDA is presented in the supplementary materials.

**References and Notes**

1. A. K. Geim, K. S. Novoselov, The rise of graphene, *Nat. Mater.* **6**, 183–191 (2007).

2. Gianluca Fiori, Francesco Bonaccorso, Giuseppe Iannaccone, Tomás Palacios, Daniel Neumaier, Alan Seabaugh, Sanjay K. Banerjee, Luigi Colombo, Electronics based on two-dimensional materials, *Nat. Nanotechnol.* **9**, 768–779 (2014).

3. Max C. Lemme, Deji Akinwande, Cedric Huyghebaert, Christoph Stampfer, 2D materials for future heterogeneous electronics, *Nat. Commun.* **13**, 1392 (2022).

4. A. K. Geim, I. V. Grigorieva, Van der Waals heterostructures, *Nature* **499**, 419–425 (2013).

5. Rebeca Ribeiro-Palau, Changjian Zhang, Kenji Watanabe, Takashi Taniguchi, James Hone, Cory R. Dean, Twistable electronics with dynamically rotatable heterostructures, *Science* **361**, 690-693 (2018).

6. S. Cahangirov, M. Topsakal, E. Aktürk, H. Şahin, S. Ciraci, Two- and One-Dimensional Honeycomb Structures of Silicon and Germanium, *Phys. Rev. Lett.* **102**, 236804 (2009).

7. Sivacarendran Balendhran, Sumeet Walia, Hussein Nili, Sharath Sriram, Madhu Bhaskaran, Elemental Analogues of Graphene: Silicene, Germanene, Stanene, and Phosphorene, *Small* **11**, 640-652 (2014).

8. Tobias G Gill, Antoine Fleurence, Ben Warner, Henning Prüser, Rainer Friedlein, Jerzy T Sadowski, Cyrus F Hirjibehedin, Yukiko Yamada-Takamura, Metallic atomically-thin layered silicon epitaxially grown on silicene/$ZrB_2$, *2D Mater.* **4**, 021015 (2017).

9. Mikhail R. Baklanov, Christoph Adelmann, Larry Zhao, Stefan De Gendt, Advanced Interconnects: Materials, Processing, and Reliability, *ECS J. Solid State Sci. Technol.* **4**, Y1-Y4 (2015).

10. Shiqian Liu, Keith Sweatman, Stuart McDonald, Kazuhiro Nogita, Ga-Based Alloys in Microelectronic Interconnects: A Review, *Materials* **11**, 1384 (2018).

11. C.-K. Hu, J.M.E. Harper, Copper interconnections and reliability, *Mater. Chem. Phys.* **52**, 5-16 (1998).

12. J. Harper, E. Colgan, C. Hu, P. Hummel, L. Buchwalter, C. Uzoh, Materials Issues in Copper Interconnections, *MRS Bulletin* **19**, 23-29 (1994).

13. W. P. Su, J. R. Schrieffer, A. J. Heeger, Solitons in Polyacetylene, *Phys. Rev. Lett*. **42**, 1698-1701 (1979).

14. T. Ozaki, Variationally optimized atomic orbitals for large-scale electronic structures, *Phys. Rev. B* **67**, 155108 (2003).

15. W. Kohn, L. J. Sham, Self-Consistent Equations Including Exchange and Correlation Effects, *Phys. Rev.* **140**, A1133-A1138 (1965).

16. D. M. Ceperley, B. J. Alder, Ground State of the Electron Gas by a Stochastic Method, *Phys. Rev. Lett.* **45**, 566–569 (1980).

17. John P. Perdew, Kieron Burke, Matthias Ernzerhof, Generalized Gradient Approximation Made Simple, *Phys. Rev. Lett.* **77**, 3865-3868 (1996).

18. P. Giannozzi, S. Baroni, N. Bonini, M. Calandra, R. Car, C. Cavazzoni, D. Ceresoli, G. L. Chiarotti, M. Cococcioni, I. Dabo, A. Dal Corso, S. Fabris, G. Fratesi, S. de Gironcoli, R. Gebauer, U. Gerstmann, C. Gougoussis, A. Kokalj, M. Lazzeri, L. Martin-Samos, N. Marzari, F. Mauri, R. Mazzarello, S. Paolini, A. Pasquarello, L. Paulatto, C. Sbraccia, S. Scandolo, G. Sclauzero, A. P. Seitsonen, A. Smogunov, P. Umari, R. M. Wentzcovitch, QUANTUM ESPRESSO: a modular and open-source software project for quantum simulations of materials, *J.Phys.: Condens. Matter* **21**, 395502 (2009).

19. P. Giannozzi, O. Andreussi, T. Brumme, O. Bunau, M. Buongiorno Nardelli, M. Calandra, R. Car, C. Cavazzoni, D. Ceresoli, M. Cococcioni, N. Colonna, I. Carnimeo, A. Dal Corso, S. de Gironcoli, P. Delugas, R. A. DiStasio Jr, A. Ferretti, A. Floris, G. Fratesi, G. Fugallo, R. Gebauer, U. Gerstmann, F. Giustino, T. Gorni, J Jia, M. Kawamura, H.-Y. Ko, A. Kokalj, E. Küçükbenli, M .Lazzeri, M. Marsili, N. Marzari, F. Mauri, N. L. Nguyen, H.-V. Nguyen, A. Otero-de-la-Roza, L. Paulatto, S. Poncé, D. Rocca, R. Sabatini, B. Santra, M. Schlipf, A. P. Seitsonen, A. Smogunov, I. Timrov, T. Thonhauser, P. Umari, N. Vast, X. Wu, S. Baroni, Advanced capabilities for materials modelling with Quantum ESPRESSO, *J.Phys.: Condens. Matter* **29**, 465901 (2017).

20. Florian Gimbert, Chi-Cheng Lee, Rainer Friedlein, Antoine Fleurence, Yukiko Yamada-Takamura, Taisuke Ozaki, Diverse forms of bonding in two-dimensional Si allotropes: Nematic orbitals in the $MoS_2$ structure, *Phys. Rev. B* **90**, 165423 (2014).

21. H. Şahin, S. Cahangirov, M. Topsakal, E. Bekaroglu, E. Akturk, R. T. Senger, S. Ciraci, Monolayer honeycomb structures of group-IV elements and III-V binary compounds: First-principles calculations, *Phys. Rev. B* **80**, 155453 (2009).

22. Andrew J. Mannix, Xiang-Feng Zhou, Brian Kiraly, Joshua D. Wood, Diego Alducin, Benjamin D. Myers, Xiaolong Liu, Brandon L. Fisher, Ulises Santiago, Jeffrey R. Guest, Miguel Jose Yacaman, Arturo Ponce, Artem R. Oganov, Mark C. Hersam, Nathan P. Guisinger, Synthesis of borophenes: Anisotropic, two-dimensional boron polymorphs, *Science* **350**, 1513-1516 (2015).

23. Yalong Jiao, Fengxian Ma, Xiaolei Zhangb, Thomas Heine, A perfect match between borophene and aluminium in the $AlB_3$ heterostructure with covalent Al–B bonds, multiple Dirac points and a high Fermi velocity, *Chem. Sci.* **13**, 1016-1022 (2022).

24. Chi-Cheng Lee, Yung-Ting Lee, Masahiro Fukuda, Taisuke Ozaki, Tight-binding calculations of optical matrix elements for conductivity using nonorthogonal atomic orbitals: Anomalous Hall conductivity in bcc Fe, *Phys. Rev. B* **98**, 115115 (2018).

25. Daniel Gall, Electron mean free path in elemental metals, *J. Appl. Phys.* **119**, 085101 (2016).

26. Mitsuaki Kawamura, FermiSurfer: Fermi-surface viewer providing multiple representation schemes, *Comput. Phys. Commun.* **239**, 197-203 (2019).

27. Choongyu Hwang, David A. Siegel, Sung-Kwan Mo, William Regan, Ariel Ismach, Yuegang Zhang, Alex Zettl, Alessandra Lanzara, Fermi velocity engineering in graphene by substrate modification, *Sci. Rep.* **2**, 590 (2012).

28. Paolo E. Trevisanutto, Christine Giorgetti, Lucia Reining, Massimo Ladisa, Valerio Olevano, Ab Initio *GW* Many-Body Effects in Graphene, *Phys. Rev. Lett.* **101**, 226405 (2008).

29. D. C. Elias, R. V. Gorbachev, A. S. Mayorov, S. V. Morozov, A. A. Zhukov, P. Blake, L. A. Ponomarenko, I. V. Grigorieva, K. S. Novoselov, F. Guinea, A. K. Geim, Dirac cones reshaped by interaction effects in suspended graphene, *Nat. Phys.* **7**, 701-704 (2011).

30. K. Momma, F. Izumi, VESTA 3 for three-dimensional visualization of crystal, volumetric and morphology data, *J. Appl. Cryst.* **44**, 1272-1276 (2011).

**Acknowledgments**

The calculations were carried out using the facilities in JAIST and Tamkang University. C.-C. L. acknowledges the Ministry of Science and Technology of Taiwan for financial support under Contract No. MOST 110-2112-M-032-016-MY2. Y.-T. L. acknowledges the Ministry of Science and Technology of Taiwan for financial support under Contract No. MOST 111-2811-M-A49-507. Y. Y.-T. acknowledges support from JSPS KAKENHI Grant Numbers 21H05232 and 21H05236.